\documentclass{article}
\input epsf

\begin{document}

\title{Concavity for nuclear bindings, thermodynamical functions and density 
functionals}

\author{B. R. Barrett \\
 Physics Department, University of Arizona, \\
Tucson, AZ 85721, USA \\
        B. G. Giraud \\
Institut de Physique Th\'eorique, DSM, CE Saclay, \\
91191 Gif-sur-Yvette, France \\
        B. K. Jennings \\
TRIUMF, Vancouver BC, V6T2A3, Canada\\
        N. P. Toberg \\
Polar Ocean Physics Group, Centre for Mathematical Sciences, \\ 
Cambridge, CB3 0WA, UK}

\date{\today} 
\maketitle

\begin{abstract}

Sequences of experimental ground-state energies for both odd and even $A$
are mapped onto concave patterns cured from convexities due to pairing and/or 
shell effects. The same patterns, completed by a list of excitation energies, 
give numerical estimates of the grand potential $\Omega(\beta,\mu)$ for a 
mixture of nuclei at low or moderate temperatures $T=\beta^{-1}$ and at many 
chemical potentials $\mu.$ The average nucleon number 
$\langle {\bf A} \rangle(\beta,\mu)$ then becomes a continuous variable, 
allowing extrapolations towards nuclear masses closer to drip lines. We 
study the possible concavity of several thermodynamical functions, such as 
the free energy and the average energy, as functions of 
$\langle {\bf A} \rangle.$ Concavity, which always occurs for the free energy 
and is usually present for the average energy, allows easy interpolations and 
extrapolations providing upper and lower bounds, respectively, to binding 
energies. Such bounds define an error bar for the prediction of binding 
energies. Finally we show how concavity and universality are related in the 
theory of the nuclear density functional.

\end{abstract}

\begin{center}
\end{center}

\section{Introduction}

The observation of a valley of stability and the search for mass formulae 
belong to the oldest subjects studied in nuclear physics \cite{Wei} and 
continue to be of great interest today \cite{Cak}. Given the
neutron and proton numbers $N$ and $Z$ as independent variables and the
corresponding atomic number, $A \equiv N+Z,$ terms such as volume energy 
$\propto A$, surface tension $\propto A^{\frac{2}{3}},$ Coulomb energy 
$\propto Z(Z-1)/A^{1/3},$ symmetry energy $\propto (N-Z)^2/A,$ {\it etc.}, 
flourish in the literature, and a great deal of attention has been dedicated 
to the consideration of finer corrections, such as, for instance, terms
$s(N,Z)$ and $p(N,Z)$ that account for shell and pairing effects,
respectively, and further correlations. This work is motivated by 
the observation that the dominant terms, such as 
$\propto A,$ $\propto Z(Z-1)$ and $\propto (N-Z)^2$, define a notoriously
{\it concave} energy surface. 

Upper and lower bounds to nuclear binding energies can be deduced from such a 
concavity, provided that deviations from concavity, possibly induced by 
subdominant terms like $\propto A^{\frac{2}{3}},$ $s(N,Z),$ $p(N,Z),$ 
{\it etc.}, can be corrected. For the sake of simplicity this paper first
considers only sequences of isotopes and, thus, takes advantage of concavity
with respect to $N$ only; $Z$ is frozen. In Section 2 we begin with a theory 
at zero temperature and show how elementary, invertible transformations of 
data can generate truly concave patterns. This is obtained by an analysis of 
the table of second differences between binding energies, then by a removal 
of the pairing energy, and finally by an {\it ad hoc}, but minimal, parabolic 
term added to the nuclear energies, if necessary.

Since concavity is also a property of several thermodynamical functions, an 
extension of the zero temperature analysis to finite temperatures is in 
order. In Section 3 we discuss properties of that grand potential, 
$\Omega(\beta,\mu),$ which can be deduced from the experimental data after 
their tuning. Other thermodynamical functions are also considered, and 
their concavity is tested. Bounds are found, and an error bar for predictions
is estimated.

A generalization to concavity with respect to both $N$ and $Z$ is briefly 
studied in Section 4. An additional motivation for our investigation of 
concavity properties is the need, in density functional theories, of 
concavity, if the universality \cite{unedf} of a density functional must be
obtained. A solution will be shown.

Finally, a discussion and conclusion are given in Section 5.

\section{Concavity for experimental ground-state energies}

Our argument is best illustrated numerically, by using a sequence of isotopic
ground-state binding energies, $-E_A.$ We choose to work with the tin isotopes
(Sn), because they provide a large number of known isotopes for testing our 
method. In addition, in Section 3, we will extend our approach to finite 
temperatures using a grand potential, which requires that a sufficiently 
large number of excited states are also known. For instance, a table of $25$ 
ground-state bindings for the tin isotopes from $^{110}$Sn to $^{134}$Sn 
reads, in keV,

\noindent
\{  934571,        942744,       953531,       961274,       971574,
    979120,        988684,       995627,   \, 1004954,   \, 1011438, 
\  1020546,    \  1026716,   \  1035529,   \  1041475,   \  1049963,
   1055696,       1063889,      1069439,      1077345,      1082676, 
   1090293,       1095540,      1102851,      1105320,      1109239\}.
Despite a standard linear trend because of a ``not too much fluctuating 
average energy per nucleon'', this list is not making a smooth pattern,
even less of a concave one. The full line in Fig. 1, where, for graphical 
convenience, we have added to each $E_A$ a constant term, $115$ MeV and a 
linear term, $7.5 \times A$ MeV, shows the amount of irregularity in the
pattern. The main source of irregularity is, obviously, the pairing effect.
If it can be removed, concavity emerges. Concavity for a sequence of isotones
is not a surprise, because of the $Z(Z-1)$ nature of the Coulomb term. For a 
sequence of isotopes, however, there is no such obviously quadratic term 
available, notwithstanding the empirical modelization of a symmetry energy 
$\propto (N-Z)^2.$ Our choice of Sn, because of its long isotope sequence, 
illustrates this concavity more dramatically.

\vspace{-2.0cm}
\begin{figure}[htb] \centering
\mbox{  \epsfysize=80mm
         \epsffile{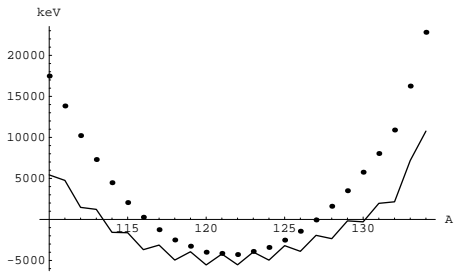}
     }
\vspace{-2.5cm}
\caption{Sn isotope energies: irregular line joining raw experimental 
ground-state energies $E_A+7500\, A+115000$; pairing and parabolic 
corrections give the nonconnected dots.} 
\end{figure}

\vspace{-2.5cm}
\begin{figure}[htb] \centering
\mbox{  \epsfysize=80mm
         \epsffile{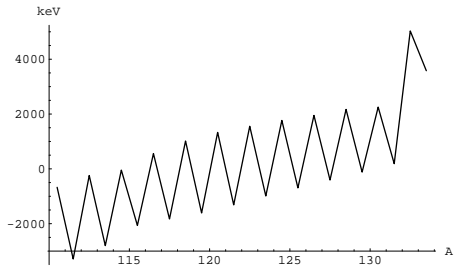}
     }
\vspace{-2.5cm}
\caption{Staggering effect, seen from first differences $D=E_{A+1}-E_A+7500$.} 
\end{figure}

\noindent
Most energies stated in this paper are in units of keV. Our data come from 
Ref. \cite{Wap03}. See also \cite{Wap95,Wap97,PE}. These sources usually quote
the binding energy per nucleon instead of the total binding energy itself and
such values per nucleon are given to varying numbers of significant figures,
from six to seven for the Sn isotopes \cite{Wap03} that we consider.
Consequently, we quote and use all our binding energies to six or seven
significant figures. Our values are generally accurate to the order of two
to three keV near the center of the sequence, and tens of keV at both ends
for the Sn isotopes. See, in particular, \cite{Wap03} for error bars.

The staggering effect is clear from the $24$ first differences, 
$E_{A+1}-E_A+7500,$

\noindent
\{-673, -3287, -243, -2800, -46, -2064,  557,  -1827,  1016,  -1608, 
  1330,  -1313,  1554,  -988,  1767, -693, 1950,  -406,  2169,  -117, 
  2253,  189,  5031,  3581\},
see Fig. 2.

The list of $23$ second differences (SDs), $SD=E_{A+1}-2 E_A+E_{A-1},$

\noindent
\{-2614, 3044, -2557, 2754, -2018, 2621, -2384,  2843, -2624, 2938, 
 -2643, 2867, -2542,  2755, -2460, 2643, -2356, 2575,  -2286, 2370, 
 -2064, 4842, -1450\},
is insensitive to the constant and linear terms we used for graphical 
convenience. It gives estimates of the ``curvatures'' of the pattern. It turns
out to be far from containing only positive numbers. The nonconnected points 
shown in Fig. 3 represent this pattern of SDs.
A systematic oscillation, reflecting the staggering effect, is found. 
Alternating signs are obviously due to the gains of binding for even Sn 
nuclei because of pairing. The oscillation between SDs centered at odd and 
even nuclei has, roughly speaking, a constant amplitude. Notice, however, 
the maximum in the list, $4842,$ due to the shell closure at $^{132}$Sn. 

\vspace{-3.0cm}
\begin{figure}[htb] \centering
\mbox{  \epsfysize=100mm
         \epsffile{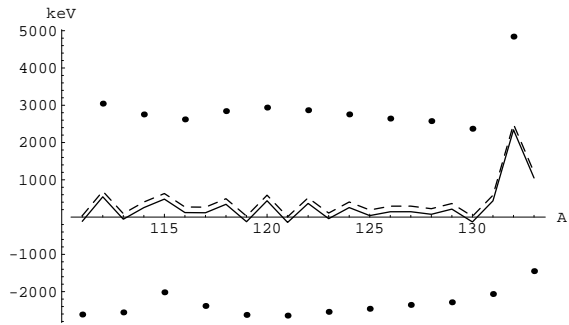}
     }
\vspace{-3cm}
\caption{Second differences $SD=E_{A+1}-2E_A+E_{A-1}$. Dots, raw data; 
solid line, result of the pairing correction; dashed line, final result 
after the parabolic correction.} 
\end{figure}

Add now to each {\it even} nucleus energy a fixed number, for example 
$p(N,Z)=1250$ keV, to tentatively suppress the increase of binding due to 
pairing. The resulting list of SDs is attenuated by an amount equal to 
$\pm 2p$, as illustrated by the full line in Fig. 3. The attenuated list reads,

\noindent
\{-114, 544, -57, 254, 482, 121, 116, 343, -124, 438, -143, 367, -42, 255, 40, 
 143, 144, 75, 214, -130, 436, 2342, 1050\}.
All numbers are now significantly smaller than their partners in the previous 
list of SDs, except for the smaller, but still large maximum at $^{132}$Sn. 
This maximum is positive, however, and causes no difficulty for concavity. 
The interesting point is rather the most negative number in the list, namely 
$-143$ keV. All negative curvatures can be converted into positive ones if 
we add to every energy an artificial, parabolic correction, 
$P \times (A-122)^2,$ with $P=75$ keV. Incidentally, the lowest point, 
$A_0=122,$ of the ``added parabola'' is arbitrary, because SDs will increase 
by just a constant, namely twice the coefficient $P$ of the $A^2$ term. After 
this $2P=150$ keV shift, the whole sequence of SDs becomes positive,

\noindent
\{36, 694, 93, 404, 632, 271, 266, 493, 26, 588, 7, 517, 108, 405, 190, 293, 
 294, 225, 364, 20, 586, 2492, 1200\},
see the dashed line shown in Fig. 3.

In short, a ``concavity ensuring'' manipulation for the isotope energies 
consists in replacing each energy $E_A$ by 
$E^{\, \prime}_A =
E_A + p \times Mod[A+1,2] + P \times (A-A_0)^2+115000+7500\, A.$ (We recall
that the terms which we have added are here just for graphical, and 
later, numerical convenience; they do not influence the theory.)
With $p=1250,$ $P=75$ and $A_0=122,$ the list of such tuned energies
$E^{\, \prime}_A$ (shown by the nonconnected dots in Fig. 1) reads,

\noindent
\{17479, 13831, 10219, 7301, 4476, 2055, 266, -1252, -2504, -3263, -3996, 
 -4141, -4279, -3900, -3413, -2521, -1439, -64, 1605, 3499, 5757, 8035, 
 10899, 16255, 22811\}.
The choice of the two parameters, p=1250 keV and P=75 keV, is empirical: one 
looks for a pairing correction leading to a modest, if not minimal, parabolic 
correction inducing concavity. Other choices for $\{p,P\}$ are possible, but,
obviously, within a small range around $1250$ and $75.$ Furthermore, such 
parameters must be readjusted for different regions of the table of nuclei,
but it is again obvious that readjustments will be moderate; for instance, 
the order of magnitude for pairing will always be around $\sim 1.0$ to 
$\sim 1.5$ MeV. Analyzing short sequences obviously leaves fewer negative SDs
to be compensated by the artificial, parabolic term, and, hence should induce
smaller values of $P.$ For this reason, we would expect short sequences to
give often better extrapolations.

Once concavity is obtained, it is straightforward that extrapolations 
from two points on the concave pattern allow predictions of lower bounds to 
nuclear energies and interpolations give upper bounds. Then, from such bounds 
for energies $E^{\, \prime},$ one recovers bounds, of strictly the same 
quality, for the physical energies $E.$ This obtains by subtracting 
from each $E^{\, \prime}$ bound its ``tuning term''. 

Assume now that $^{110}$Sn were unknown and one had done a brute force 
extrapolation, $2 E_{111}-E_{112}=-931957.$ Compared with $E_{110}=-934571,$
this prediction underbinds $^{110}$Sn by $2614$ keV. Consider rather 
$2 E^{\, \prime}_{111}-E^{\, \prime}_{112}=17443,$ to be compared with 
$E^{\, \prime}_{110}=17479.$ A slight overbinding, by 36 keV, is found. 
Although this small error is likely accidental, it is clear that systematic 
lower bounds will be found. Naturally, once a value for $E^{\, \prime}_A$ is 
predicted, one recovers as good an estimate for $E_A$ after removing the 
``concavity manipulation terms'', which are known explicitly.
 In the present case, the tuning of the data added $952050$ to $E_{110}.$ 
The same $952050$ must be subtracted from that value, 
$2 E^{\, \prime}_{111}-E^{\, \prime}_{112}=17443,$ extrapolated from the 
concave pattern, yielding a final result of $-934607,$ to be compared with 
$E_{110}=-934571,$ which obviously exhibits the same slight overbinding, 
$36$ keV. Based on the data in \cite{Wap03}, the experimentally measured
energy for $^{110}$Sn has an error bar $\pm 14$ keV; hence, in this 
case, the experimental error bar and the uncertainty in our theoretical value 
have the same order of magnitude.

If $^{132}$Sn were unknown, the brute force extrapolation gives, 
$2 E_{131}-E_{130}=-1100787,$ to be compared with $E_{132}=-1102851,$ 
showing an underbinding equal to $2064$ keV, clearly failing to reproduce 
the shell closure effect. Now, from the concave pattern, we obtain 
$2 E^{\, \prime}_{131}-E^{\, \prime}_{130}=10313,$ to be compared with 
$E^{\, \prime}_{132}=10899,$ producing an overbinding by $586$ keV. The 
same overbinding is obviously found if one subtracts 
from $2 E^{\, \prime}_{131}-E^{\, \prime}_{130}=10313,$ the
tuning difference, $E^{\, \prime}_{132}-E_{132}=1113750,$ with the result, 
$-1103437,$ to be compared with  $E_{132}= -1102851.$  See Fig. 4, which 
also illustrates an extrapolation using $E_{133}$ and $E_{134}$ and an 
interpolation using $E_{131}$ and $E_{133}$.

\vspace{-1.0cm}
\begin{figure}[htb] \centering
\mbox{  \epsfysize=75mm
         \epsffile{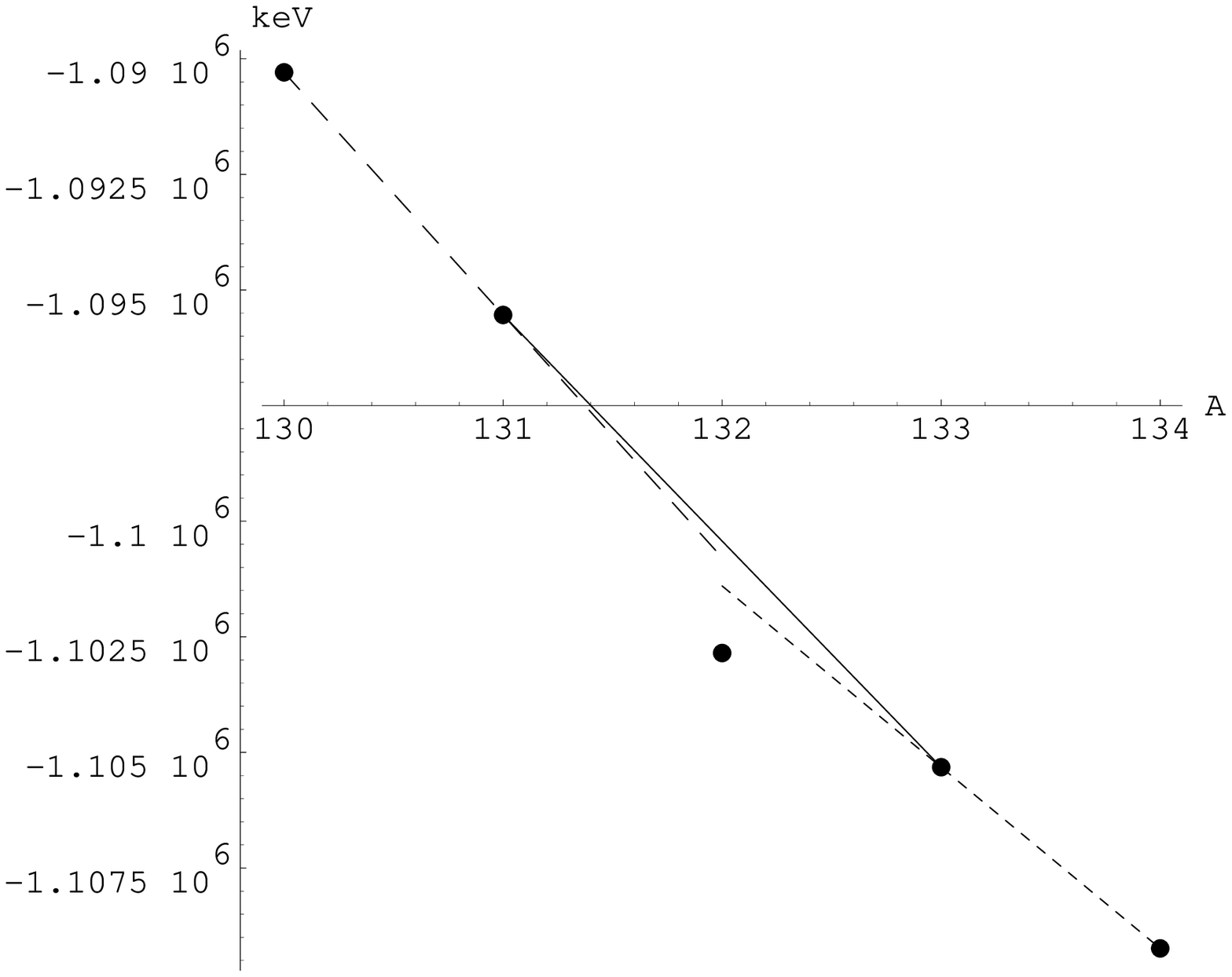}
     }
\mbox{  \epsfysize=75mm
         \epsffile{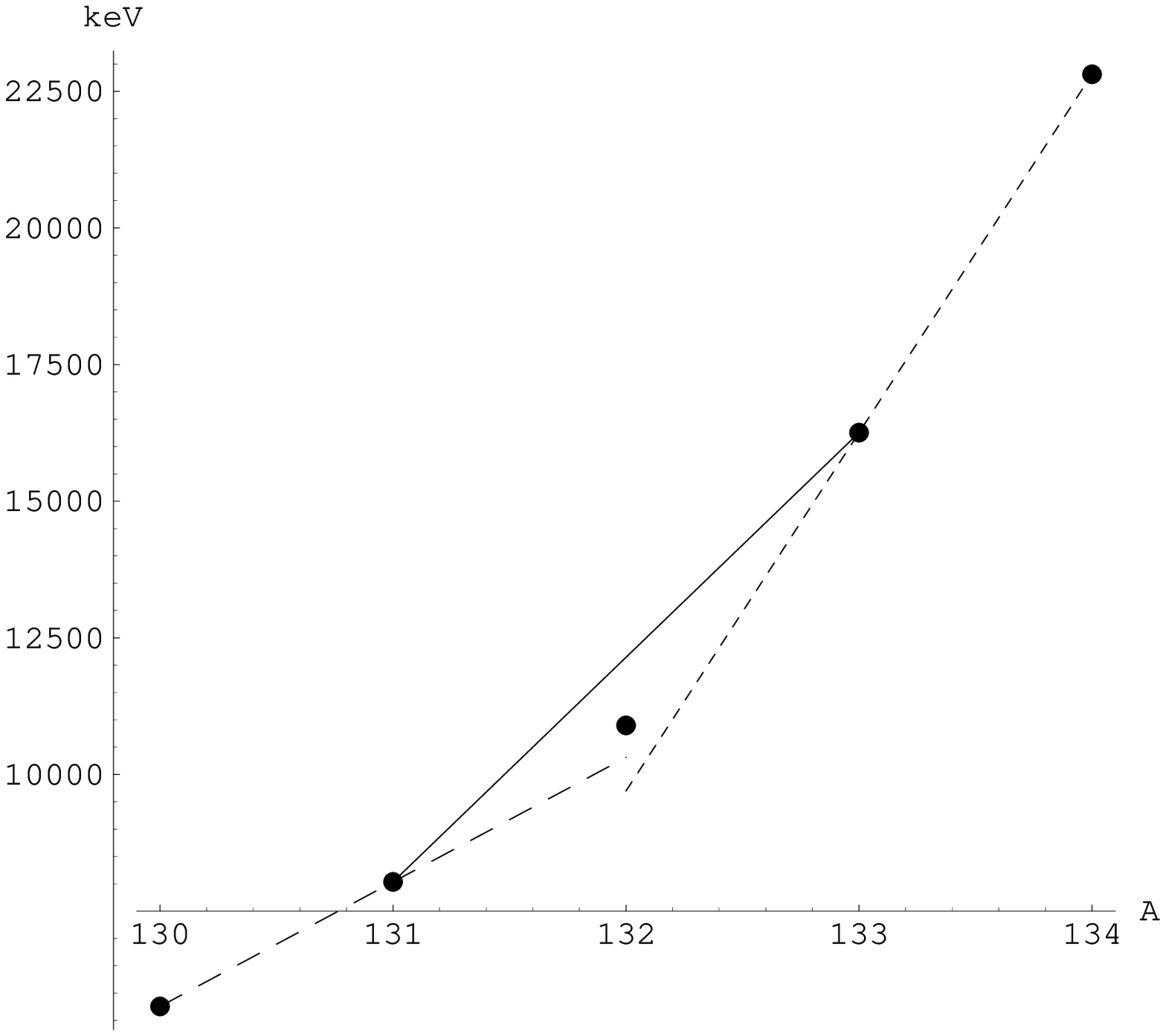}
     }
\vspace{-1.5cm}
\caption{Left, raw energies: full line, interpolation from $^{131}$Sn and
$^{133}$Sn; long dashes, extrapolation from  $^{130}$Sn and $^{131}$Sn;
short dashes, extrapolation from  $^{134}$Sn and $^{133}$Sn. Dots represent
the experimental energies $E_A.$ Right: same inter- and extrapolations,
concave data.} 
\end{figure} 

Let us now also compare for an odd rather
than even isotope, {\it e.g.}, $^{117}$Sn, two extrapolations and one 
interpolation from raw data and concave data, respectively.
With our choice of parameters, the tuning difference,
$E^{\, \prime}_{117}-E_{117},$ is $994375,$ with $E_{117}= -995627.$ 
Extrapolations from lower masses
give, $2 E_{116}-E_{115}=-998248,$ and 
$2 E^{\, \prime}_{116}-E^{\, \prime}_{115}-994375=-995898,$
hence, overbindings of $2671$ and $271,$ respectively. Extrapolations from
higher masses give, $2 E_{118}-E_{119}=-998470,$ and 
$2 E^{\, \prime}_{118}-E^{\, \prime}_{119}-994375=-996120,$ leading to 
overbindings of $2843$ and $493,$ 
respectively. Interpolations from the nearest nuclei give, 
$(E_{116}+E_{118})/2=-996819,$ and 
$(E^{\, \prime}_{116}+E^{\, \prime}_{118})/2-994375=-995494,$ with 
now overbinding of $1192$ and underbinding of $133,$ respectively. The poor 
results from raw data stress the importance of pairing corrections. 
The much better results from the concave pattern do take this pairing into 
account, but, furthermore, they again make it trivial that 
extrapolations give overbindings, while interpolations give underbindings.

For the sake of completeness, we now consider a case with a two-step 
extrapolation. If the tuned pattern followed a perfect parabolic form, 
$E^{\, \prime}_A=a\, A^2+b\, A+c,$ the formula, 
$E^{\,\prime}_{A+2} - 3\, E^{\, \prime}_A + 2\, E^{\, \prime}_{A-1} = 6\, a,$
predicts, for the lower bound, an error which is a little larger than
desirable. This is because of the coefficient, $6,$ and the fact that the 
``curvature parameter'', $a,$ has here an order of magnitude in the few 
hundred keV range, see those residual second differences used for the dashed
curve in Fig. 3. This indicates that further stages of the theory, with
polynoms at least, are worthy of consideration. However, to remain within the
``linear'' scope of the present paper, we give the results obtained for
$^{135}$Sn, $-1113208,$ and $^{136}$Sn, $-1115027,$ when we use the lower
bound estimates from the formula,
$3\, E^{\, \prime}_A - 2\, E^{\, \prime}_{A-1}.$ From Ref. \cite{Wap03}, the
predictions, which are based on neighbor nuclei with the same $Z$ and $N$
parities, are $-1111320$ and $-1115064,$ respectively, with error
bars of $\sim \pm .5$ MeV. We do find a lower bound for $^{135}$Sn, with
apparently a severe, but mostly expected, error of a couple of MeVs.
For $^{136}$Sn, we apparently fail to generate a lower bound, but by so small
a difference that our results stay well inside the error bars. 

To summarize Section 2, when concavity occurs, one concludes 
that extrapolations from two points on the concave pattern allow predictions 
of lower bounds to nuclear energies. In the same way,
interpolations provide upper bounds. The quality of such ``concavity bounds'' 
from zero temperature data is found to be good; interpolations and 
extrapolations from the raw, nonconcave pattern, are clearly less 
satisfactory. But there is a more profound reason why a concave pattern is 
necessary. Indeed, several thermodynamical functions, governed by theorems 
proving their concavity, have a notoriously singular limit at zero 
temperature: they become nonanalytical and are just piecewise continuous. 
Their limit plots are made of segments; derivatives are discontinuous at 
turning points. Because of the staggering effect, the {\it concave envelope} 
of the raw pattern of $E_A$'s would contain only the even isotopes. Concavity 
is, thus, {\it necessary} for a theory that must accommodate both odd and 
even nuclei. Therefore, the thermodynamical functions studied in Section 3 
preferably use concave energies $E^{\, \prime}_A$ and the corresponding 
excited state energies $E^{\, \prime}_{nA}.$

\section{Concavity with thermodynamical functions}

Consider the particle number operator ${\bf A}$ and a familiar nuclear 
Hamiltonian ${\bf H}=\sum_{i=1}^A t_i + \sum_{i>j=1}^A v_{ij},$ where $A,$ 
$t$ and $v$ are the mass number, one-body kinetic energy and two-body 
interaction, respectively. Nuclear data tables \cite{istps,nndc} give precise 
values for a large number of lowest-lying eigenvalues $E_{nA}$ of ${\bf H},$
for many nuclei. One may, thus, reasonably estimate the grand partition 
function,
\begin{equation}
{\cal Z}(\mu,\beta) = {\rm Tr}\, 
\exp\left[ \beta \left( \mu {\bf A} - {\bf H} \right)  \right] =
\sum_{nA} (2 j_{nA}+1) 
\exp\left[ \beta \left( \mu A - E_{nA} \right)  \right],
\label{definZ}
\end{equation} 
provided that 
i) the temperature, $T=\beta^{-1},$ is low enough to allow a truncation
of the spectrum to include only those states provided by the tables and 
ii) the chemical potential, $\mu,$ selects mainly those nuclei in which we
are interested. Let $\langle \ \rangle$ denote, as usual, a statistical
average. The (equilibrium!) density operator in Fock space,
$\rho={\cal Z}^{-1} \exp\left[ \beta\, ( \mu {\bf A} - {\bf H} ) \right],$
ensures that the following grand potential,  
$\Omega(\mu,\beta) = \langle\, (  {\bf H} - \mu {\bf A}  )\, \rangle - 
T\, {\bf S},$ is minimal in the space of many-body density matrices 
with unit trace, since, by definition,
$\langle {\bf A} \rangle = {\rm Tr}\, \rho\, {\bf A},$ 
$\langle {\bf H} \rangle = {\rm Tr}\, \rho\, {\bf H},$ with the entropy,
${\bf S}=                 -{\rm Tr}\, (\rho\, \log \rho).$

We shall rather use $(-\Omega)$ in the following, to make upcoming proofs of 
concavity slightly easier. This grand potential also reads,
\begin{equation}
-\Omega(\mu,\beta) = \beta^{-1} \ln\, {\cal Z} = 
\beta^{-1} \ln \left\{ \sum_{nA} (2 j_{nA}+1) 
\exp\left[ \beta \left( \mu A - E_{nA} \right)  \right] \right\},
\end{equation}

Simple manipulations then give the relevant statistical averages 
$\langle \ \rangle$ of particle numbers and energies, together with their 
derivatives and fluctuations, 
\begin{equation}
\partial (-\Omega) / \partial \mu = \langle {\bf A} \rangle =  \sum_A A\ p_A,
\ \ \ p_A= {\cal Z}^{-1} \sum_n (2 j_{nA}+1) 
\exp\left[ \beta \left( \mu A - E_{nA} \right)  \right],
\label{avrgnmb} 
\end{equation}
and
\begin{equation}
\partial^2 (-\Omega) / \partial \mu^2 =
\partial \langle {\bf A} \rangle / \partial \mu = \beta\, 
\left(\, \langle {\bf A}^2 \rangle - \langle {\bf A} \rangle^2\, \right),
\label{flucA}
\end{equation}
then
\begin{equation}
\partial (-\Omega) / \partial T = {\bf S} = \ln {\cal Z} - 
\beta\, \langle\, (\mu {\bf A} - {\bf H})\, \rangle,
\label{ntrp}
\end{equation}
or as well,
\begin{equation}
\langle {\bf H} \rangle = {\cal Z}^{-1} \sum_{nA} (2 j_{nA}+1) 
E_{nA}\, \exp\left[ \beta \left(\mu A - E_{nA} \right)  \right].
\label{avrgnrg}
\end{equation}
Furthermore,
\begin{equation}
\partial^2 (-\Omega) / \partial T^2 = \beta^3 \left[
\langle\, \left( \mu {\bf A} - {\bf H} \right)^2 \, \rangle -
\langle\, \left( \mu {\bf A} - {\bf H} \right)   \, \rangle^2 \right],
\label{flucB} 
\end{equation}
and
\begin{equation}
\partial^2 (-\Omega) / (\partial \mu\, \partial T) = -\beta^2 \left[
\langle\, {\bf A} \left( \mu {\bf A} - {\bf H} \right) \, \rangle -
\langle {\bf A} \rangle \,
\langle\, \left( \mu {\bf A} - {\bf H} \right)   \, \rangle \right].
\label{crossAB} 
\end{equation}
For our investigations, we will freeze $\beta$ as real and consider 
functions of a real $\mu.$ It is then well known that $(-\Omega)$ is
a concave function of $\mu$ and $T.$ In turn, the double Legendre transform,
with respect to both $\mu$ and $T,$
\begin{equation}
\mu\ \partial (-\Omega) / \partial \mu + T\, \partial (-\Omega) / \partial T + 
\Omega = \langle {\bf H} \rangle,
\end{equation}
shows that $\langle {\bf H} \rangle$ is a concave function of both 
$\langle {\bf A} \rangle$ and ${\bf S},$ the conjugate variables of $\mu$ 
and $T,$ respectively.

In the following, we do not perform the full, double Legendre transform.  We 
rather retain an intermediate representation, with $\langle {\bf A} \rangle$ 
and either $T$ or $\beta.$ We stay with real variables and functions.  We 
stress that, while ${\bf A}$ has a discrete spectrum,
$\langle {\bf A} \rangle$ is continuous, a monotonically increasing function
of $\mu,$ smooth provided $\beta$ is finite. The monotonicity results from
Eq. (\ref{flucA}). Actually, at low temperatures, strong derivatives signal
the onset of discrete jumps due to the integer spectrum of ${\bf A},$ but we
may stay away from this ``jumpy'' regime in the following, at least 
temporarily. Anyhow, at any fixed, finite $\beta,$ the smoothness and 
monotonicity of $\langle {\bf A} \rangle$ with respect to $\mu$ allows a 
reasonably easy numerical calculation of the inverse function
$\mu\left( \langle {\bf A} \rangle \right).$  Thus, a main argument of this 
Section is that at fixed temperatures we will use  $\langle {\bf A} \rangle$ 
as a continuous variable and attempt extrapolations towards unknown nuclei.

For this, given a value of $T,$ we keep track of $\langle {\bf A} \rangle$ 
and $\langle {\bf H} \rangle$ as functions of $\mu.$ Since the functional 
inversion from $\langle {\bf A} \rangle (\mu)$ to 
$\mu(\langle {\bf A} \rangle)$ is reasonably easy, we can plot 
$\langle {\bf H} \rangle$ in terms of $\langle {\bf A} \rangle$ and
attempt an extrapolation for further values of $\langle {\bf A} \rangle.$
This extrapolation can be considered as a ``candidate'' for a mass formula, 
at that finite temperature $T.$

According to Eq. (\ref{flucB}), the average, constrained energy,
$\langle \left({\bf H} - \mu {\bf A}\right) \rangle,$ is a monotonically
decreasing function of $\beta.$ Furthermore, at least for negative chemical 
potentials $\mu$, and, more generally, if $A$ has an upper bound, the operator,
${\bf H} - \mu {\bf A},$ is bounded from below. Therefore, there is a
convergence of the process consisting in
i) extrapolating with respect to $\mu$ both 
$\langle \left({\bf H} - \mu {\bf A}\right) \rangle$ 
and $\langle {\bf A} \rangle$ for fixed values of $\beta,$ then
ii) eliminating $\mu$ to generate the $\beta$-parametrized ``mass formula'' 
$\langle {\bf H} \rangle \left(\langle {\bf A} \rangle, \beta \right),$ 
and finally iii) considering the limit of this mass formula when 
$\beta \rightarrow +\infty.$ Alternately, it is equivalent, and maybe more
efficient, to first eliminate $\mu$ and then extrapolate the ``mass formula''
$\langle {\bf H} \rangle \left(\langle {\bf A} \rangle, \beta \right),$ first
with respect to $\langle {\bf A} \rangle,$ then with respect to $\beta.$

Is there concavity in this intermediate representation?  Clearly, a simple 
Legendre transform of $(-\Omega),$ with respect to $\mu$ only, returns a 
free energy, ${\bf F}=\langle {\bf H} \rangle - T\, {\bf S},$ as a concave 
function of $\langle {\bf A} \rangle$ and $T.$ If $T$ is low enough to allow 
the product $T\, {\bf S}$ to be neglected, then, at fixed $T,$ one may accept
that $\langle {\bf H} \rangle$ is an ``almost'' concave function of 
$\langle {\bf A} \rangle.$ This assumption will be tested by the numerical 
results which follow. Incidentally, a straightforward calculation of
$A'' \left( \langle {\bf A} \rangle, \beta \right)
\equiv \partial^2 \langle {\bf H} \rangle / 
\left(\partial \langle {\bf A} \rangle \right)^2$ yields,
$
A'' \propto
\left( \langle {\bf A}^2 \rangle - \langle {\bf A}\rangle^2 \right) 
\langle {\bf A}^2 {\bf H} \rangle + 
\left( \langle {\bf A} \rangle \langle {\bf A}^2 \rangle - 
\langle {\bf A}^3 \rangle \right) 
\langle {\bf A} {\bf H} \rangle + 
\left( \langle {\bf A} \rangle \langle {\bf A}^3 \rangle - 
  \langle {\bf A}^2 \rangle^2 \right) 
\langle {\bf H} \rangle,
$
with a positive factor,
$\left( \langle {\bf A}^2 \rangle - \langle {\bf A} \rangle^2 \right)^{-3}.$
More simply,
\begin{equation}
\frac{\partial^2 \langle {\bf H} \rangle} {\partial \langle {\bf A} \rangle^2}
\left( \langle {\bf A} \rangle, \beta \right)\, \propto\,
\langle\, (\Delta {\bf A})^2\, \rangle\ 
\langle\, (\Delta {\bf A})^2\ \Delta {\bf H}\, \rangle -
\langle\, (\Delta {\bf A})^3\, \rangle\ 
\langle\, \Delta {\bf A}\     \Delta {\bf H}\, \rangle, 
\label{secderisimpl}
\end{equation}
if one uses the centered operators, 
$\Delta {\bf A}={\bf A} - \langle {\bf A} \rangle$ and 
$\Delta {\bf H}={\bf H}-\langle {\bf H} \rangle.$ From 
Eq. (\ref{secderisimpl}), concavity for $\langle {\bf H} \rangle$ is unclear; 
we shall have to test it numerically.

\vspace{-1.0cm}
\begin{figure}[htb] \centering
\mbox{  \epsfysize=80mm
         \epsffile{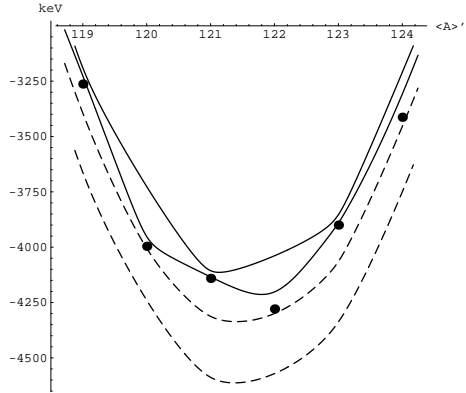}
     }
\vspace{-1.5cm}
\caption{Energy $\langle {\bf H} \rangle'$ (full curves) and free energy
${\bf F}'$ (dashed ones) as functions of $\langle {\bf A} \rangle',$
calculated from concavity tuned energies $E^{\, \prime}_{nA}.$ Dots represent
tuned, ground-state energies,
$E^{\, \prime}_{0A} \equiv E^{\, \prime}_A,$ already used in Fig. 1 and the
rhs part of Fig. 4. Upper full and lower dashed curves, $T=150$ keV. Lower
full and upper dashed curves, $T=60$ keV. Notice how the full curves turn out
to be concave.}
\end{figure}

The results, which follow, use \cite{istps,nndc} for the 
excited-state energies of the Sn isotopes.
Using the first $10$ levels of $^{110}$Sn to $^{132}$Sn, hence a maximum
excitation of, typically, $\sim 1.2$ MeV for odd isotopes and twice as much
for even ones, we calculate ${\cal Z},$ see Eq. (\ref{definZ}). For those
rare cases, where the spin $j_{nA}$ is ambiguous, we choose the lowest
of the suggested spins. If the spin is completely unknown, we set it to be 
either $0$ or $1/2$, according to $A.$ These tactics minimize the statistical 
influence of such rare cases. The highest excited levels that we use have an 
excitation energy of order a few MeV; hence, we restrict our thermodynamical 
analysis to temperatures of order $\sim 500$ keV at most. It should be noted 
that we use the concavity tuned energies, 
$E^{\, \prime}_{nA}=E_{nA}+115000+7500\, A+1250\, Mod[A+1,2]+75\, (A-122)^2.$ 
Consequently, all our calculated quantities, $\langle {\bf A} \rangle',$  
${\bf S}',$ $\langle {\bf H} \rangle',$ {\it etc.}, derived from 
Eqs. (\ref{avrgnmb}), (\ref{ntrp}), and (\ref{avrgnrg}), for instance, carry
a prime superscript. We want to stress that the spurious exponential factor
created by the constant, $115000,$ appears in the same way in numerators and
denominators and, hence, cancels out; it only helps the graphics and numerics.
Similarly, the linear term, $7500\, A,$ means but a change of reference for
the chemical potential. Only the pairing and parabolic tuning terms make the
primed quantities truly distinct from those ``raw'', unprimed ones, obtained
with the untuned energies $E_{nA}.$  

As an example of our results we choose for Fig. 5 the sequence of six isotopes 
from $^{119}$Sn to $^{124}$Sn, because of its still severely contrasted list
of second differences, $\{588, 7, 517, 108\},$ after tuning.  This produces a
serious geometrical constraint on the plots of $\langle {\bf H} \rangle'$ and 
${\bf F}'$, as they reach their zero temperature limit, and better validates 
our approach. The upper curves in Fig. 5 are the plots of the function 
$\langle {\bf H} \rangle'(\langle {\bf A} \rangle')$ when $T=60$ keV (lower 
full curve) and $T=150$ keV (upper full curve), respectively.  The increase of 
$\langle {\bf H} \rangle',$ when $T$ increases is transparent. The striking 
result is the apparent concavity of both curves. It is also found that the
lower temperature, $60$ keV, is low enough to allow $\langle {\bf H} \rangle'$
to run, in practice, almost through the ground-state energies of both even and
odd nuclei.  As a check of our results, we repeated our calculations using
the untuned energies, {i.e.,} those which lack concavity. We then found that
the low temperature limit of the $\langle {\bf H} \rangle$ curve went through
the {\it even} nuclei only, namely, the limit curve follows the {\it concave
envelope} of the experimental pattern. 

For graphical and pedagogical convenience, Fig. 5 shows the plots for
six nuclei only, but the same observations hold for full plots, with 
$110 \le \langle {\bf A} \rangle' \le 134.$ As a test, we also calculated 
$\langle {\bf H} \rangle'$ and its low temperature limit when the levels of 
$^{132}$Sn are omitted from the trace sum, Eq. (\ref{definZ}).  In that case,
the concave envelope goes through the dots representing the two odd nuclei, 
$^{131}$Sn and $^{133}$Sn, and ignores the dot representing $^{132}$Sn. 
Similar verifications of other concave envelopes were obtained by removing 
other nuclei.

To verify whether concave envelopes and concavity for 
$\langle {\bf H} \rangle'$ result from negligible values of the entropy term
in the free energy, or, more precisely, negligible values of its second 
derivative, a calculation of $T\, {\bf S}'$ is in order.  Figure 6 shows the 
difference, $T\, {\bf S}',$ between the energy and the free energy, 
as a function of $\langle {\bf A} \rangle',$ for $T=60$ and $150$ keV, 
respectively. At the higher temperature, $150$ keV, wiggling effects seem 
to be small enough to allow for the concavity of $\langle {\bf H} \rangle'.$
At the lower temperature, $60$ keV, the wiggling is stronger.  As a
consequence of such a strong wiggling, possibly translating into strong
second derivatives, it is not excluded that SDs coming from $T\, {\bf S}'$
might prevent $\langle {\bf H} \rangle'$ from having the proven concavity
property of the free energy. We must, therefore, numerically calculate second
derivatives, see Eq. (\ref{secderisimpl}).

\vspace{-2.0cm}
\begin{figure}[htb] \centering
\mbox{  \epsfysize=75mm
         \epsffile{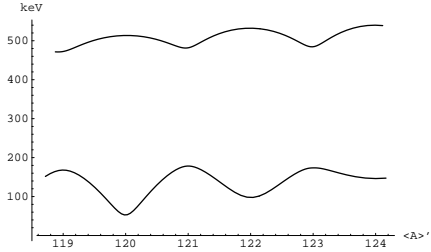}
     }
\vspace{-2.5cm}
\caption{Entropy energy $T\, {\bf S}' (\langle {\bf A} \rangle').$ Upper
(lower) curve, $T=150$ ($60$) keV.}
\end{figure}

In Fig. 7 we show, for tuned data, plots of the rhs of 
Eq. (\ref{secderisimpl}), divided by 
$\left( \langle {\bf A}^2 \rangle' - \langle {\bf A} \rangle'^2 \right)^3.$
This quantity must remain positive if $\langle {\bf H} \rangle'$ shows
concavity. The full line represents the situation when $T=60$ keV, the dashed
line corresponds to $T=150$ keV. We see that the second derivative,
$\partial^2 \langle {\bf H} \rangle' / \partial \langle {\bf A} \rangle'^2,$
remains positive almost always. For low temperatures, however, negative
values may appear. For instance, the full curve, corresponding to
$T=60$ keV, indicates small, but definitely negative values around
$^{119}$Sn. Numerical tests, which are not easy because of a difference 
effect between the two terms present in Eq. (\ref{secderisimpl}),
show that the occurrence of such negative, actually moderate, values for 
lower temperatures might be somewhat frequent, while not systematic. 
Furthermore, such ``negativity accidents'' turn out to be worse when we use 
untuned data, maybe because  the untuned data lack concavity in the first
place.

\vspace{-1.5cm}
\begin{figure}[htb] \centering
\mbox{  \epsfysize=90mm
         \epsffile{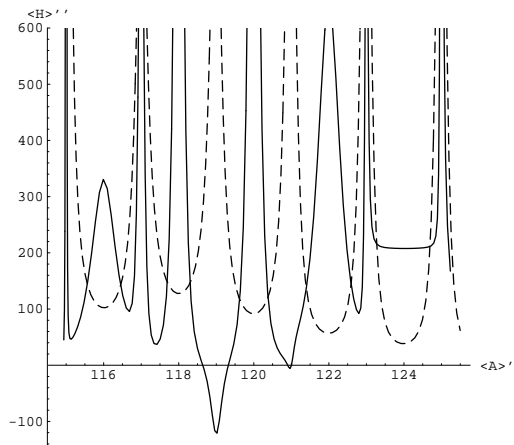}
     }
\vspace{-2.0cm}
\caption{(a) Behavior of $H'' =
\partial^2 \langle {\bf H} \rangle' / \partial \langle {\bf A} \rangle'^2;$
full (dashed) curve, $T=60$ ($150$)keV. Notice tiny excursions in 
negative value territory by the low $T$ curve.}
\end{figure}

A likely reason for the negativity accidents with tuned data might be that 
the fluctuation of ${\bf A}$ is not large enough to justify our use of 
$\langle {\bf A} \rangle'$ as a continuous variable. Since it interpolates 
between integers, a fluctuation of order $\sim 1,$ or at least $\sim .5,$ 
might be necessary. As shown by the plots in Fig. 8, corresponding to $T=60$ 
keV and $150$ keV for the lower and upper curves, respectively, a minimum 
temperature of order $\sim 150$ keV is needed to avoid too small a fluctuation 
of the particle number. Indeed, for, e.g. $\sim 100$ keV, we found that 
the fluctuation is smaller than $.5$ for almost half of the calculated
values of $\langle {\bf A} \rangle'.$

At this stage, the situation can be summarized as follows. On the one hand, 
the tuned pattern of experimental energies shows concavity, but the concavity
of $\langle {\bf H} \rangle'$ as a function of $\langle {\bf A} \rangle'$ is 
not sure, although it seems to occur most of the time. On the other hand, we 
have a theorem proving concavity for the free energy, 
either ${\bf F} \equiv \langle {\bf H} \rangle - T\, S = \mu\, 
\langle {\bf A} \rangle + \Omega $
or ${\bf F}' \equiv \langle {\bf H} \rangle' - T\, S' = \mu\, 
\langle {\bf A} \rangle' + \Omega', $ as functions of $T$ and 
$\langle {\bf A} \rangle$ or $\langle {\bf A} \rangle',$ respectively.
For instance, elementary derivations show that, in that representation where
$\langle {\bf A} \rangle$ (or $\langle {\bf A} \rangle'$) and $\beta$ are the 
primary variables, 
\begin{equation}
\frac{\partial{\bf F}'}{\partial \beta} = \beta^{-2}\, S',\ \ \ \  
\frac{\partial{\bf F}'}{\partial \langle {\bf A} \rangle'} = \mu\ \ \ \  
{\rm and}\ \ \ \ 
\frac{\partial^2{\bf F}'}{\partial \langle {\bf A} \rangle'^2} = \frac{T}
{ \langle {\bf A}^2 \rangle' - \langle {\bf A} \rangle'^2 }\, .
\end{equation}
To summarize this discussion, we see that the removal of the entropy 
term, leading from the free energy to just the energy, can sometimes destroy 
the concavity depending upon the temperature, but only weakly, {\it e.g.,} 
see Fig. 7. 

\vspace{-1.5cm}
\begin{figure}[htb] \centering
\mbox{  \epsfysize=75mm
         \epsffile{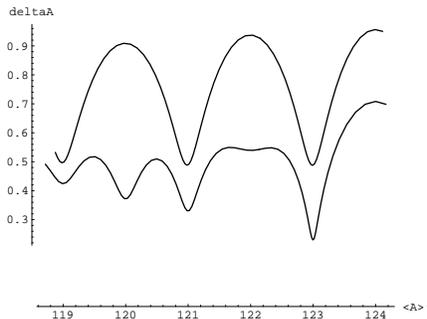}
     }
\vspace{-2.0cm}
\caption{Particle number fluctuation. Upper (lower)  curve, $T=150$ ($60$)
keV.}
\end{figure}

A compromise may be found so that $T$ is low enough to make $T {\bf S}'$ small
with respect to ${\bf F}'$ and high enough to ensure 
both sufficient values of $\Delta {\bf A}$ and a tolerable oscillating of 
$T {\bf S}'.$ Our rationale will be that $\langle {\bf H} \rangle',$ even
though it might deviate from concavity, will stay close enough to the concave
${\bf F}'.$ Their difference, $T {\bf S}',$ a positive quantity, will define
an error bar between a lower bound ${\bf F}'$ and an upper bound 
$\langle {\bf H} \rangle'$ for ground-state tuned energies.
Consider again Fig. 5. There is no need to stress how, given $T,$ the energy 
and free energy curves make a band defining upper and lower bounds for the 
experimental energies. Moreover, a similar, but narrower band is obtained if
$T$ decreases. Figs. 5, 6 and 8 suggests $T \sim 150$ keV for an error bar 
$\sim 500$ keV between energy and free energy, a smooth enough entropy energy
and a sufficient fluctuation of the particle number. This order of magnitude
for $T$ is compatible with the average level splitting, of order 
$\sim 250$ keV, that is observed from the first 10 levels of 
$^{110}$Sn, ... ,$^{132}$Sn used in our calculations.

The following properties, i) the average energy and the free
energy are increasing and decreasing, respectively, functions of increasing 
$T,$ ii) the energy is larger than the free energy and iii) the entropy term 
by which they differ vanishes when $T$ vanishes, are not big surprises. It 
can be concluded that, in so far as thermodynamical functions can be
calculated at low enough temperatures, precise ``accuracy bands'' may be
available. Their, hopefully analytic, continuation for higher and/or lower
values of $A$ than those available in nuclear tables provides a prediction
scheme for exotic nuclei.

An estimate of the entropy term is now useful. Given $\mu$ and a {\it large}
$\beta,$ let $A_0$ and $E^{\, \prime}_0$ correspond to that nucleus whose 
ground-state energy maximizes the exponential, 
$\exp[\beta\, (\mu A - E^{\, \prime}_{0A})].$
Consider now the first subdominant exponential. It might be generated by the 
first excited state of the same nucleus, or by the ground-state of one of 
its neighbors. Let $A_1$ and $E^{\, \prime}_1$ be its parameters and define 
$\Delta=\mu\, (A_1-A_0)-(E^{\, \prime}_1-E^{\, \prime}_0).$ Concavity 
guarantees that $\Delta < 0.$
Whenever $\beta$ is large enough, it is trivial to reduce the grand canonical
ensemble to a two state ensemble, and the entropy then boils down to 
$s=- e^{\beta\, \Delta}\, \beta\, \Delta.$ Hence, the product,
$T\, s =- e^{\beta\, \Delta}\, \Delta,$ vanishes exponentially fast when 
$\beta \rightarrow \infty.$ The rate of decrease is governed by that scale 
defined by $\Delta,$ to be extracted from the tuned data. Then one can 
estimate an order of magnitude for the difference between the free energy
and the energy. This estimate can be viewed as an error bar for the 
prediction of exotic nuclei via the present ``concavity method''.

We have applied the approach to several other isotopic regions, such as 
Sm, Hg, Pb. In all cases, making the switch to the concave shape improves 
our ability to make extrapolations and interpolations for unknown bindings. 
However, estimating  the free energy and the average value of ${\bf H},$
to set narrow upper and lower bounds on these binding energies, remains more 
difficult in drip line regions, because of the lack of experimental data on 
excited states in such regions. Very simple linear extrapolations still
provide, nonetheless, error bars with an order of magnitude
$\Delta E \sim \pm 400$ keV, similar to the best ``band'' seen in Fig. 5.
A few examples of the kind of results that we can obtain from this
thermodynamical error-bar analysis, when using the method described in
Sec. 2, are these conservative estimates of the bindings of $^{171}$Hg,
$^{208}$Hg and $^{209}$Hg, namely $1314.33$ MeV $\pm 400$ keV, 
$1629.13$ MeV $\pm 400$ keV and $1632.16$ MeV $\pm 400$ keV, respectively.

\section{Two-dimensional analysis; density functional}

We now want to consider binding-energy systematics in two dimensions,
{\it i.e.,} N and Z.  The second differences studied in this Section are, 
$SD_n=E_{N+1,Z}+E_{N-1,Z}-2\, E_{NZ},$ $SD_p=E_{N,Z+1}+E_{N,Z-1}-2\, E_{NZ},$ 
$SD_b=(E_{N+1,Z+1}+E_{N-1,Z-1})/2-E_{NZ}$ and 
$SD_s=(E_{N+1,Z-1}+E_{N-1,Z+1})/2-E_{NZ}.$ If $N$ and $Z$ were continuous,
and, furthermore, if $E_{NZ}$ were a smooth function of $N$ and $Z,$
the numbers, $SD_n,$ $SD_p,$ $SD_b$ and $SD_s,$ might be interpreted as 
estimates of second derivatives of the function $E(N,Z)$ in the neutron, 
proton, constant $N-Z$ and constant $A$ directions, respectively. Their 
connections with neutron, proton and p-n pairings and with a symmetry energy
are also transparent.

Setting aside what we consider to be ``light'' nuclei, namely those of the
s-, p-, sd- and f$_{7/2}$ shells, we calculated these four SDs for the rest
of the known nuclear table. Their maxima and minima are found to be of order 

\noindent
$\sim \{4840,6180,4530,5900,-3670,-3080,-3160,-2150\}$ keV,  respectively. 
The lhs parts of Figs. 9-12 show, in scattered plots with respect to
$A,$ the patterns of those SD values. The negative SDs, contradicting
concavity, are due mainly to pairing effects. An extension of the correction
done in Sec. 2 is in order. Let $p$ be a mass dependent \cite{DasGal}
estimate of pairing. Diminish the binding of doubly even nuclei by $p$,
increase the binding of doubly odd ones by $p,$ and leave odd nuclei
untouched. We set, empirically, $p=1160-A\, .$ There is no doubt that better
parametrizations of $p$ are possible \cite{DasGal}, but the present linear
decrease suffices for our demonstration. With this correction, the worst
SDs become

\noindent 
$\sim \{2790,3970,2470,3680,-1500,-930,-980,-120\}$ keV, respectively. As 
expected, this is a large reduction of both positive and negative amplitudes. 
This is shown by the rhs parts of Figs. 9-12, respectively.

\vspace{-2.0cm}
\begin{figure}[htb] \centering
\mbox{  \epsfysize=75mm
         \epsffile{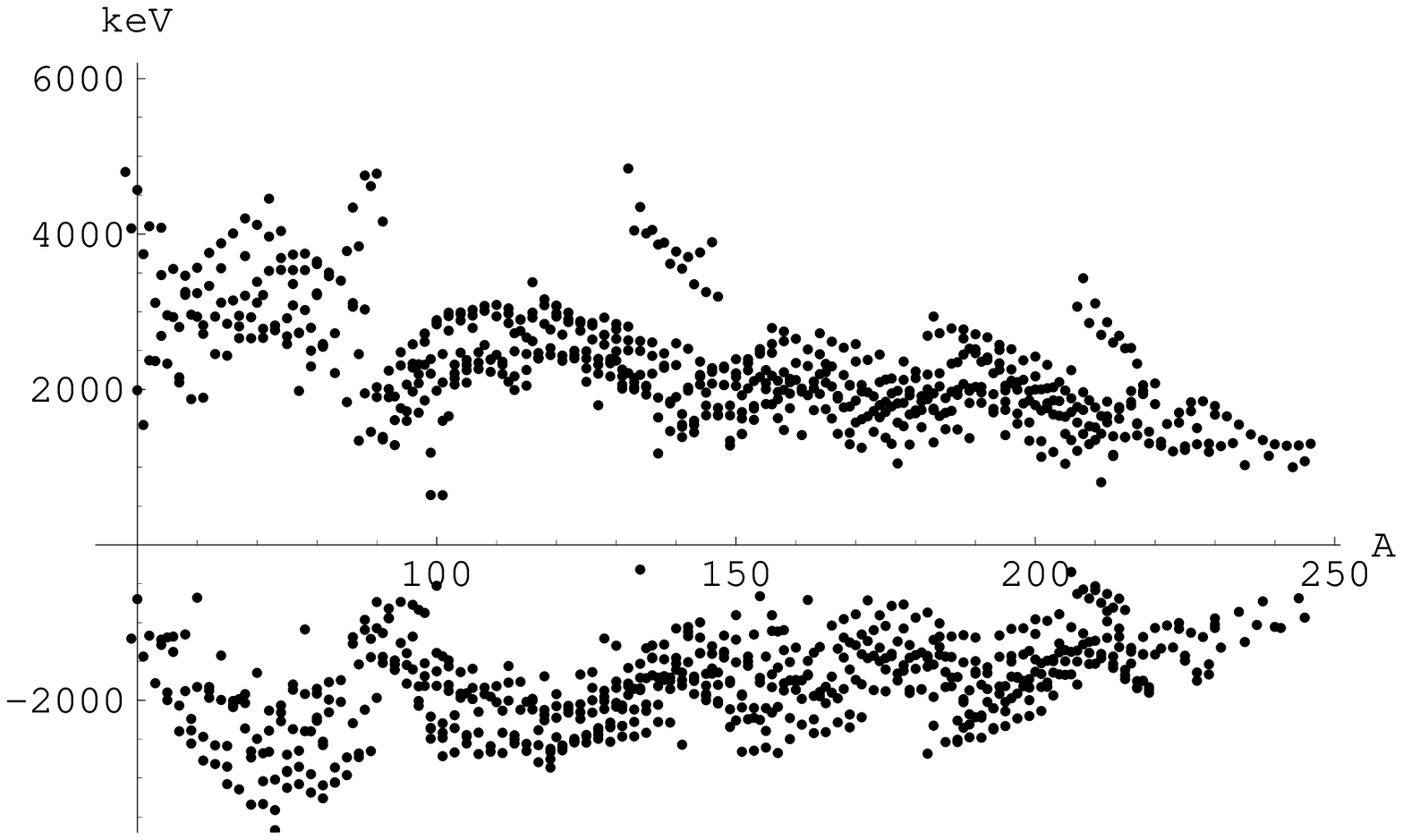}
     }
\mbox{  \epsfysize=75mm
         \epsffile{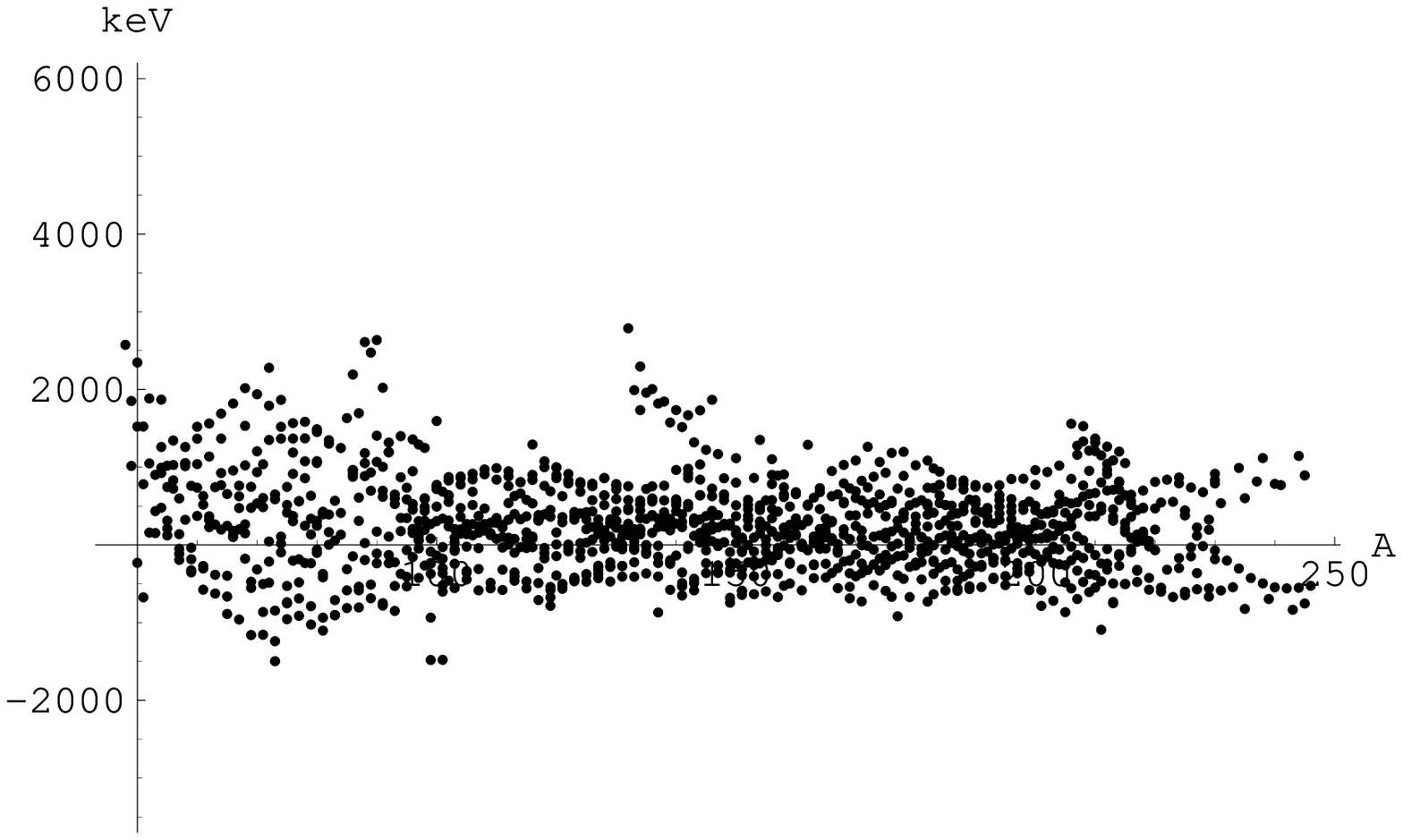}
     }
\vspace{-2.0cm}
\caption{Scatter plot of convex second differences $SD_n$ in the neutron 
direction throughout the table of known nuclei for $27 < A < 251.$
Left: with bare energies. Right: after a pairing correction.} 
\end{figure}

\vspace{-2.0cm}
\begin{figure}[htb] \centering
\mbox{  \epsfysize=75mm
         \epsffile{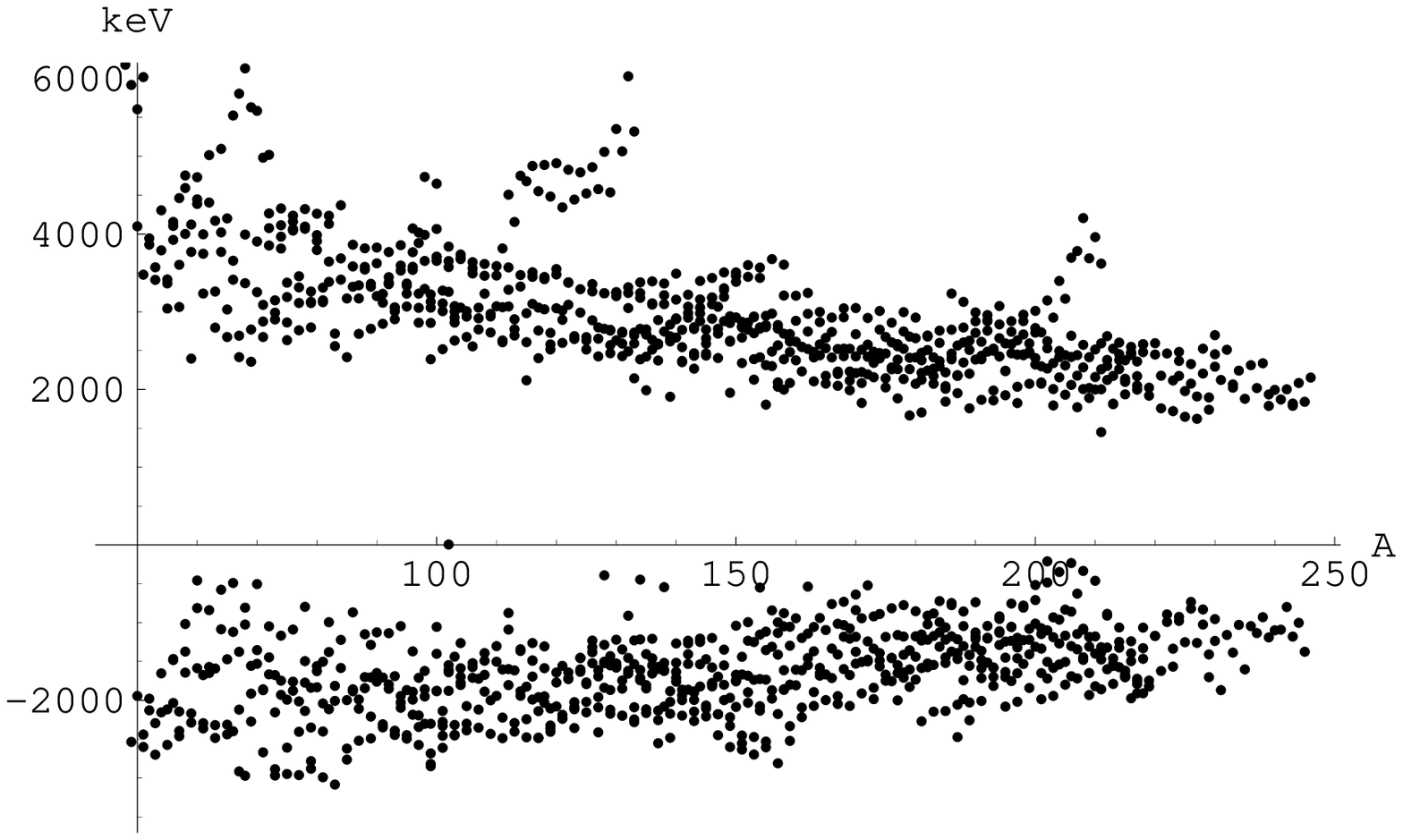}
     }
\mbox{  \epsfysize=75mm
         \epsffile{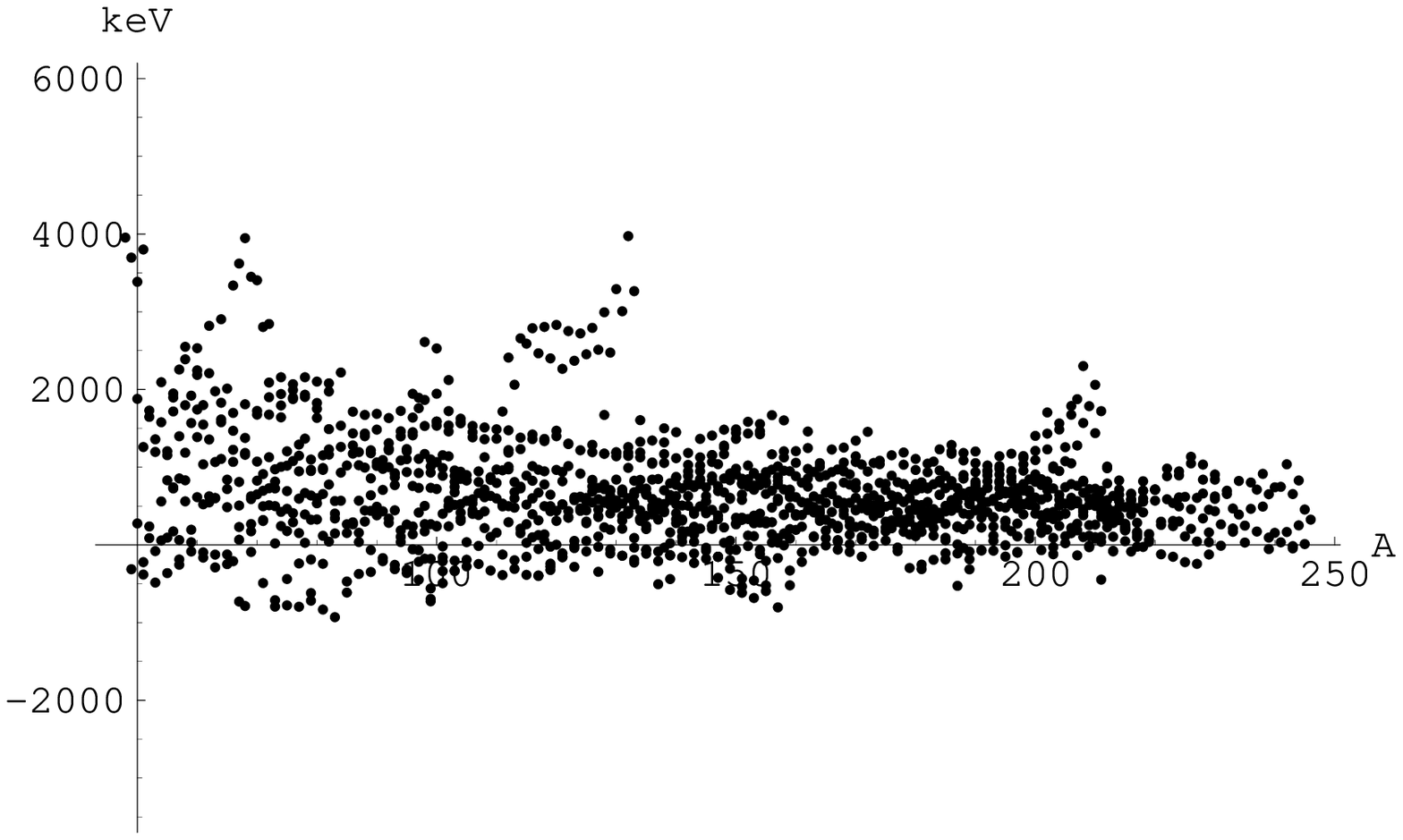}
     }
\vspace{-2.0cm}
\caption{Same as Fig. 9, but for $SD_p$ in the proton direction.} 
\end{figure}

While the pairing correction has not completely suppressed negative SDs, 
note the worst minimum is still of order  $\sim - 1500$ keV, concavity has 
been improved significantly and extrapolations are likely made more secure.

Similar to our one-dimensional analysis in Sec. 2, one can now add 
to the energies a paraboloid term, of the form, 
$u\, (N-N_0)^2 + 2v\, (N-N_0)\, (Z-Z_0) + w\, (Z-Z_0)^2,$ in order to get rid 
of the residual, negative SDs. (We recall that SDs do not depend on $N_0$ and 
$Z_0,$ which are only parameters for graphical and/or numerical convenience.) 
Notice from Fig. 12, and the corresponding worse minimum, of order 
$\sim -120$ keV, that the direction where $A$ is constant does not deviate 
much from concavity. The dominant term in the paraboloid must, therefore,
indicate a direction approximately orthogonal, and also take into account
the bend of the global stability valley towards neutron excess.

\vspace{-2.0cm}
\begin{figure}[htb] \centering
\mbox{  \epsfysize=75mm
         \epsffile{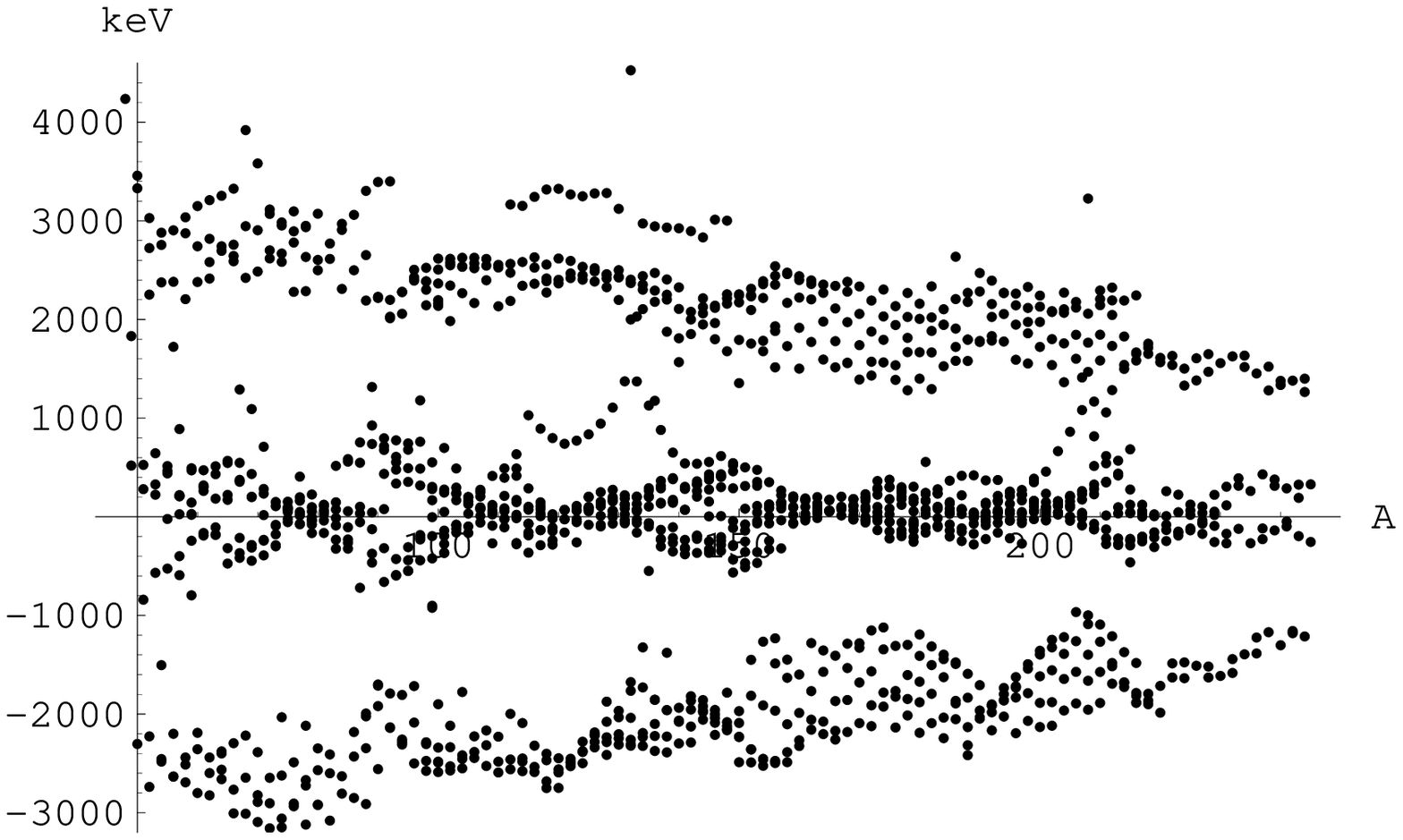}
     }
\mbox{  \epsfysize=75mm
         \epsffile{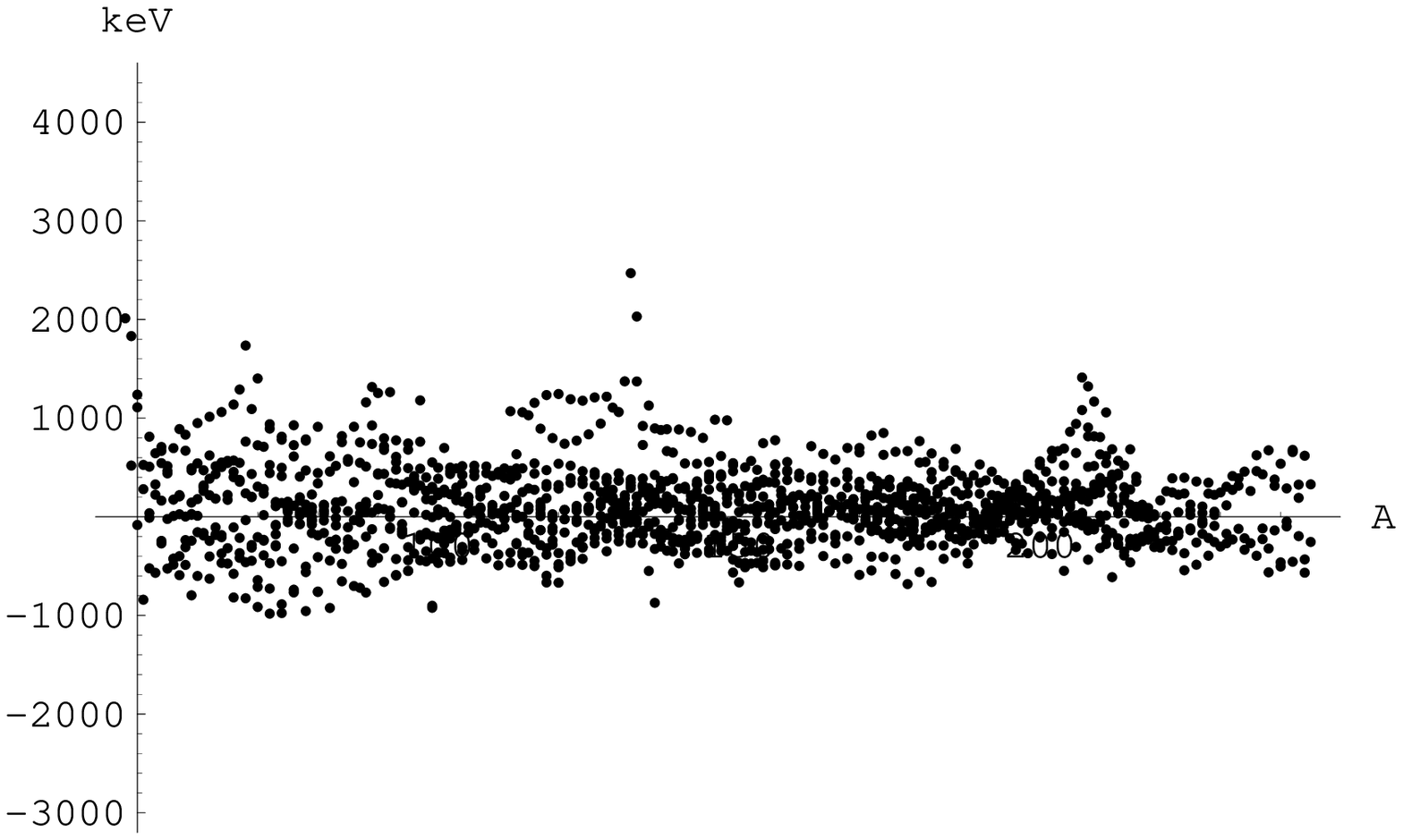}
     }
\vspace{-2.0cm}
\caption{Same as Fig. 9, but for $SD_b$ in the constant $N-Z$  direction.} 
\end{figure}

\vspace{-2.0cm}
\begin{figure}[htb] \centering
\mbox{  \epsfysize=75mm
         \epsffile{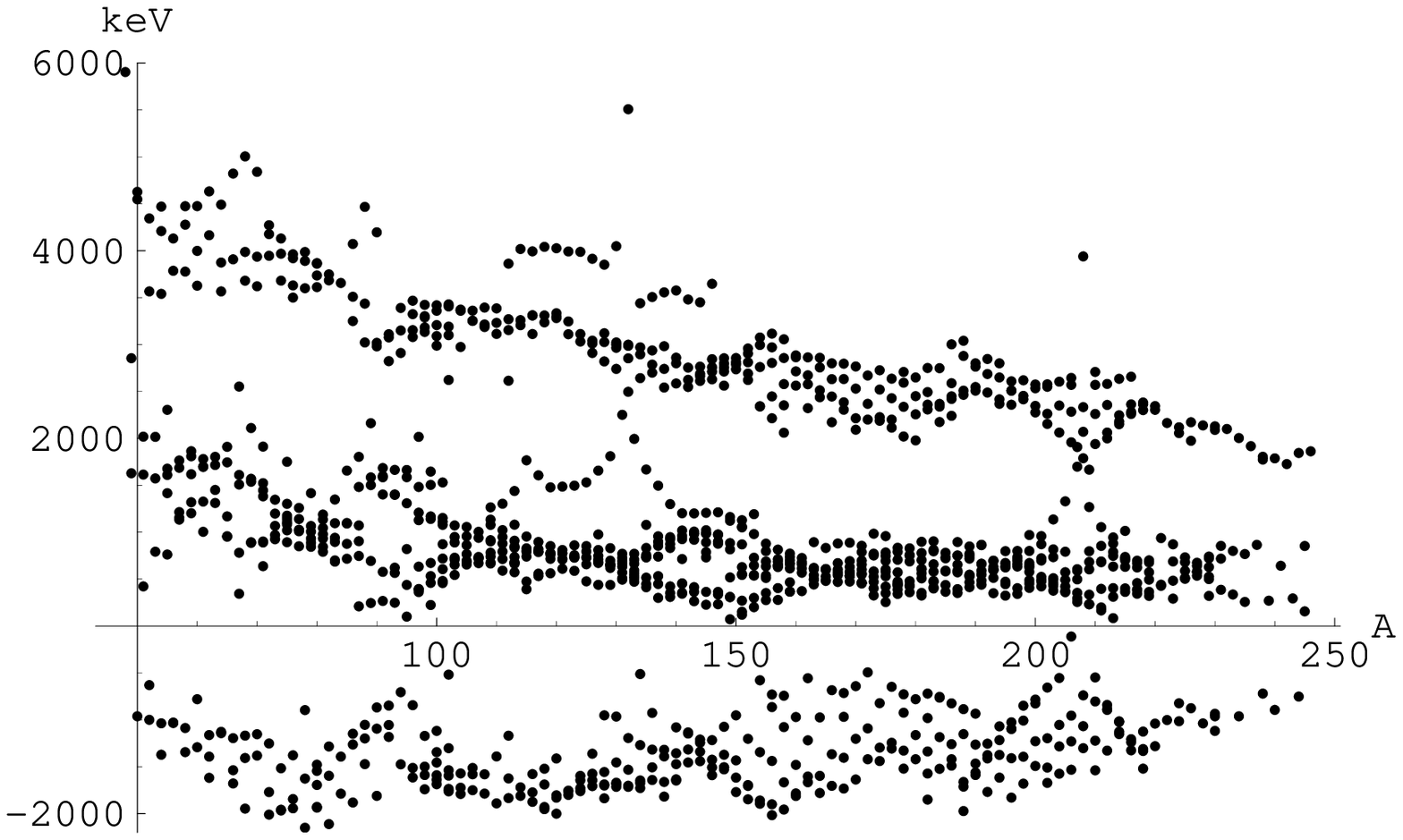}
     }
\mbox{  \epsfysize=75mm
         \epsffile{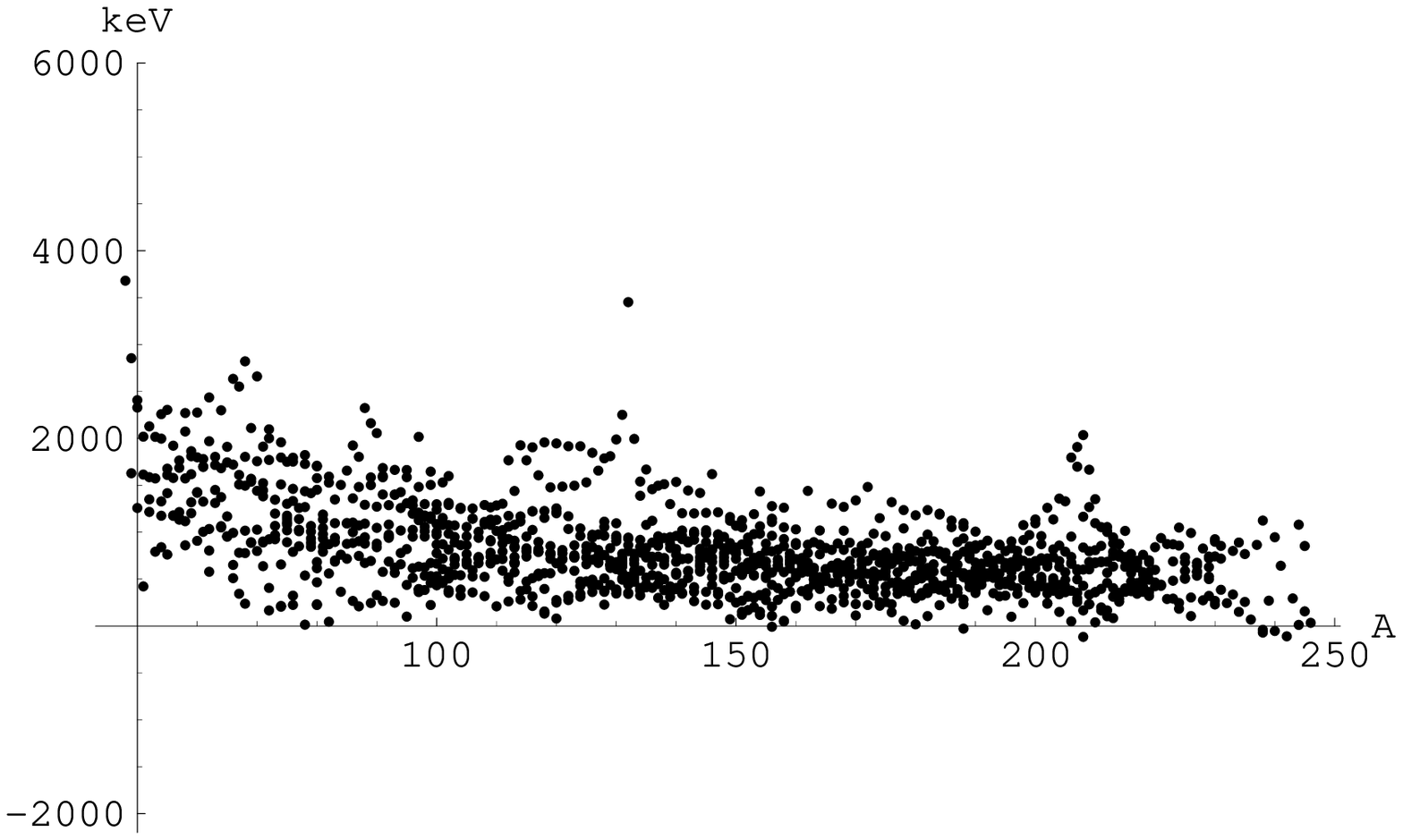}
     }
\vspace{-2.0cm}
\caption{Same as Fig. 9, but for $SD_s$ in the direction where $A$ is 
constant.} 
\end{figure}

There is, furthermore, no doubt that the two-dimensional free energy, now a 
function of both $N$ and $Z$ made continuous by a grand canonical ensemble,
remains a concave function, available for lower bounds. This is a theorem, 
which can be proven in a straightforward manner following the previous 
derivation for one dimension. Simultaneously, it is likely that the mean
energy will be again ``almost concave'', to yield upper bounds, in which
case the same error bar that was estimated in Sec. 3 remains valid.

A systematic use of such properties throughout the nuclear table is way 
beyond the scope of the present paper. Rather, we want to discuss the 
relation of concavity with the derivation of a nuclear density functional.
Recall that, given an arbitrary many-body density matrix ${\cal M}$ in Fock 
space, the density functional derives from the definition \cite{LevLie},
\begin{equation}
{\cal F}[\rho_p,\rho_n]={\rm Inf}_{{\cal M} \rightarrow \rho_p,\rho_n} 
{\rm Tr}\, {\bf H}\, {\cal M}.
\end{equation}
Here the symbol, ${\cal M} \rightarrow \rho_n,\rho_p,$ means that a 
minimization of the energy is performed upon density matrices having 
the same neutron and proton profiles $\rho_n,\rho_p.$ 
A nucleus is identified by the integrals, $N=\int d\vec r\, \rho_n(\vec r)$
and $Z=\int d\vec r\, \rho_p(\vec r),$ and a later minimization is performed 
with respect to the profiles $\rho_n,\rho_p$ under such identification 
constraints,
\begin{equation}
E_{ZN}={\rm Inf}_{\rho_n \rightarrow N,\, \rho_p \rightarrow Z}\, 
{\cal F}[\rho_n,\rho_p].
\end{equation}
The functional ${\cal F}$ should be universal, in the nuclear physics sense. 
Namely, it should not depend on $N$ and $Z.$ (In atomic and molecular physics, 
universality of the DF has a different meaning; it refers to arbitrary 
external potentials.) While BCS and Hartree-Bogoliubov calculations allow a 
distinction between even and odd particle numbers, universality in density 
profile space does not allow such a separate treatment. Observe, furthermore, 
that, because of the fact that there are many convexities in the pattern of
raw energies, there exist many cases where, for instance, three nuclei, 
$A_1,A_2,A_3,$ are such that $N_2=(N_1+N_3)/2$ and 
$Z_2=(Z_1+Z_3)/2$ and $(E_1+E_3)/2 < E_2.$ Then, in a search for $E_2,$ the 
mixture density matrix, ${\cal M}_{mix}=\left(\, | A_1 \rangle \langle A_1 | + 
| A_3 \rangle \langle A_3 | \right)/2,$
provides the correct average particle numbers $N_2,Z_2,$ but an absurd energy, 
lower than $E_2.$ Concavity is a {\it mandatory condition} for the 
universality of ${\cal F}.$

One way to simultaneously implement pairing corrections and use a minimal
 paraboloid term consists in adding to the physical ${\bf H}$ terms such as, 
$p \times [\cos^2(\pi\, {\bf N}/2) + \cos^2(\pi\, {\bf Z}/2)-1] +
u\, {\bf N}^2+ 2v\, {\bf N}\, {\bf Z} + w\, {\bf Z}^2.$ (Notice
how the cosines take advantage of the continuous nature of 
$\langle {\bf N} \rangle$ and $\langle {\bf Z} \rangle$ at finite 
temperatures.) Concavity is achieved 
if $u,v,w$ are tuned to compensate for the worst tuned SD minima listed above, 
namely $\{-1500,-930,-980,-120\}.$ A rough solution consists in taking
$p \sim 1100$ and a ``circular'' ansatz, 
$P \times \left( {\bf N}^2 + {\bf Z}^2 \right),$ with $P \sim 800,$
but less brutal solutions are obviously worthy of consideration, in order 
to minimize this tampering with the Hamiltonian. In any case, we stress
that the operators, ${\bf N}$ and ${\bf Z},$ commute with ${\bf H}$, and,
therefore, that the present counterterms do not perturb nuclear dynamics.
Once again we emphasize that this approach {\it unifies} the treatment of
odd and even nuclei. The fact that the nuclear density functional needs to
originate from a Hamiltonian, completed by such counterterms, illustrates
how important concavity is for making predictions for nuclei far from
stability, {\it i.e.,} for the so-called exotic nuclei.

\section{Summary, discussion and conclusion}

We have demonstrated how a list of ground-state energies for a sequence of
isotopes can be turned into a concave pattern. This involves simple
manipulations; for instance, an explicit term, accounting for pairing in
even nuclei, can be subtracted from the bindings. This unifies the treatment
of odd and even nuclei, a notoriously difficult problem.  If needed, a small
quadratic correction can also be added to guarantee concavity at all points
along the sequence.

Similar arguments leading to concavity clearly hold for isotones as 
well, and, furthermore, for any other sequence of neighboring nuclei in any 
direction across the nuclear table. Once this empirical tuning has 
been implemented, linear (or more general) extra- and interpolations of the 
concave pattern can provide surprisingly accurate and robust estimates of, 
or bounds for, binding energies. These tuning terms, which are added to 
induce concavity, are, of course, subtracted {\it in fine}.

This work then defined a more ambitious extra- and interpolation scheme, 
involving thermodynamical functions from a grand canonical ensemble, because 
such functions may have rigorous concavity properties. Theorems are, indeed,
available to prove such properties. 
For instance, the free energy is a concave function of the average 
particle number and is also a decreasing function of the temperature. 
We also found strong numerical evidences concerning the average energy. This 
average energy at nonzero temperature turns out, in general, to be a concave
function of the average particle number, except for ``minor accidents,''
which depend upon the temperature. 

For every given, finite temperature, we found that the average energy and 
the free energy, as functions of the average particle number, give upper and 
lower bounds, respectively, for the concave envelope of the ground-state 
energies. When the temperature vanishes, both bounds converge to the exact 
results. At this vanishing temperature, however, the analyticity of such 
thermodynamical functions is lost, because their limit is only piecewise 
continuous. It is, therefore, necessary to retain a minimum temperature 
if one wants to obtain practical extrapolations for the prediction of exotic 
nuclei.  A minimum amount of particle number fluctuation is necessary to 
validate the conversion of particle number, an initially discrete quantity, 
into a continuous variable.

We, therefore, implemented numerical estimates of several thermodynamical 
functions at moderate temperatures, a few hundred keV at most. This yields
a first result, namely a ``band'', enclosing ground-state energies between
the average energy and the free energy. The width of the band defines an error
bar which can be trusted when extrapolations are made.

A difficulty arises, however, because of an insufficient number of excited 
states for nuclei at both ends of any sequence of isotopes.  These 
excited-state energies are essential for calculating a meaningful value
of the average energy for any value of A.  More often, only the ground-state 
energy is known for such neutron very rich or neutron very poor nuclei. The 
calculation of the average and free energies is, thus, possible only in an 
interval smaller than the interval of masses where ground states are known.
Two tactics are then available, namely i) an extrapolation only of the
sequence of ground-state energies and ii) an extrapolation of the
thermodynamical functions, starting from a smaller interval. We found that
the first tactic, namely at zero temperature, is reliable, especially when
completed by the error bar derived from the finite temperature
considerations. This work also tested the second tactic, with some
success but with the limitation due to the lack of known excited 
states. The lack leads to edge effects in the calculation of thermodynamical 
functions. Therefore, if linear, or polynomial, or more general analytical
fits of such thermodynamical functions are attempted, it is better to
restrict them from a slightly smaller mass interval. Then, extrapolations
should be reliable within one and maybe two mass units towards drip lines.

We can make the strong conclusion that the combination of concavity and 
extrapolations of thermodynamical functions gives a systematic set of upper 
and lower bounds for the prediction of ground-state energies.

Our last, and perhaps most important result, is the connection between 
concavity and the universality of the density functional. In a two-dimensional
analysis, we showed that a paraboloid term with respect to the proton and
neutron number operators {\it must} be added to the Hamiltonian, so as to
guarantee obtaining consistent energy minima everywhere in density space.
This term can be made minimal if counterterms for pairing are also added.

\bigskip
{\it Acknowledgements}: We thank I. Allison for helpful discussions and 
assistance with the management of the data sets.  It is a pleasure for B. R. B.
and B. G. G. to thank TRIUMF, Vancouver, B.~C., Canada, where part of this
work was done, for its hospitality. The Natural Science and Engineering
Research Council of Canada is thanked for financial support. TRIUMF receives
federal funding via a contribution agreement through the National Research
Council of Canada. B. R. B. also thanks Institut de Physique Th\'eorique,
Saclay, France, for its hospitality, where another part of this work was
carried out, and the Gesellschaft f\"ur Schwerionenforschung (GSI),
Darmstadt, Germany, for its hospitality during the preparation of this
manuscript, and acknowledges partial support from NSF grant PHY0555396
 and from the Alexander von Humboldt Stiftung.

\end{document}